\documentclass[prl,preprint,showpacs,preprintnumbers,amsmath,amssymb]{revtex4}
\usepackage{graphicx}
\newcommand{\beq}{\begin{equation}}
\newcommand{\eeq}{\end{equation}}
\newcommand{\beqa}{\begin{eqnarray}}
\newcommand{\eeqa}{\end{eqnarray}}

\newcommand{\Sigs}{\Sigma_{\mathrm s} }
\newcommand{\Sigv}{\Sigma_{\mathrm v} }
\newcommand{\Sigo}{\Sigma_{\mathrm o} }
\newcommand{\kf}{k_{\mathrm F} }

\newcommand{\bfgamma}{\mbox{\boldmath$\gamma$\unboldmath}}
\newcommand{\veck}{\textbf{k}}

\begin{document}
\preprint{}
\title{Effective Nucleon Masses in Symmetric and Asymmetric Nuclear Matter.}
\author{E. N. E. van Dalen}
\author{C. Fuchs}
\author{Amand Faessler}
\affiliation{Institut
f$\ddot{\textrm{u}}$r Theoretische Physik, Universit$\ddot{\textrm{a}}$t
T$\ddot{\textrm{u}}$bingen,
Auf der Morgenstelle 14, D-72076 T$\ddot{\textrm{u}}$bingen, Germany}
\begin{abstract}
The momentum and isospin dependence of the in-medium nucleon mass are studied. 
Two definitions of the effective mass, 
i.e. the Dirac mass $m^*_D$ and the nonrelativistic mass $m^*_{NR}$ which parameterizes the energy spectrum,  
are compared. Both masses are 
determined from relativistic Dirac-Brueckner-Hartree-Fock 
calculations. The nonrelativistic mass shows a distinct 
peak around the Fermi momentum. The proton-neutron mass splitting in isospin asymmetric matter is 
$m^*_{D,n} <m^*_{D,p}$ and opposite for the nonrelativistic mass, i.e. $m^*_{NR,n} >m^*_{NR,p}$, which is consistent 
with nonrelativistic approaches. 
 \end{abstract}
\pacs{21.65.+f,21.60.-n,21.30.-x,24.10.Cn}
\keywords{Effective mass, Symmetric and Asymmetric Nuclear matter, Relativistic Brueckner approach}
\maketitle
The introduction of an effective mass is a common concept to characterize the 
quasi-particle properties of a particle inside a strongly interacting 
medium. It is also a well established fact that the effective nucleon mass 
in nuclear matter or finite nuclei deviates substantially from 
its vacuum value \cite{brown63,jeukenne76,mahaux85}. However, 
there exist different definitions of the effective nucleon mass which are 
often compared and sometimes even mixed up: 
the nonrelativistic effective mass $m^*_{NR}$ and the 
relativistic Dirac mass $m^*_{D}$.
These two definitions  are based on completely different physical concepts. 
The nonrelativistic mass parameterizes
the momentum dependence of the single particle potential. 
It is the result of a quadratic parameterization
of the single particle spectrum. On the other hand, the relativistic Dirac mass 
is defined through the scalar part of the nucleon self-energy in the 
Dirac field equation which is absorbed into the effective mass 
$m^*_{D} =M + \Re  \Sigma_s(k, \kf)$. This Dirac mass is a smooth function of 
the momentum. In contrast, the nonrelativistic effective 
mass - as a model independent result -  
shows a narrow enhancement near the Fermi surface due to an enhanced 
level density \cite{brown63,jeukenne76,mahaux85}.

Although related, these different definitions of the effective mass have 
to be used with care when relativistic and nonrelativistic approaches are 
compared on the basis of effective masses. While the Dirac 
mass is a genuine relativistic quantity the nonrelativistic 
mass  $m^*_{NR}$ can be determined 
from both, relativistic as well as nonrelativistic approaches. 
A heavily discussed topic is in this context the proton-neutron mass 
splitting in isospin asymmetric nuclear matter. This question is of 
importance for the forthcoming new generation of radioactive beam 
facilities which are devoted to the investigation of the isospin 
dependence of the nuclear forces at its extremes. However, presently 
the predictions for the 
isospin dependence of the effective masses differ substantially \cite{baran04}. 

Brueckner-Hartree-Fock (BHF) calculations \cite{zuo99,muether04} 
predict a proton-neutron mass splitting of $m^*_{NR,n} > m^*_{NR,p}$ in 
isospin asymmetric nuclear matter. 
This stands in contrast to relativistic mean-field (RMF) theory. When only a 
vector isovector $\rho$-meson is included Dirac phenomenology 
predicts equal masses $m^*_{D,n}= m^*_{D,p}$ while the inclusion of the 
scalar isovector $\delta$-meson, i.e. $\rho+\delta$, leads to 
$m^*_{D,n} < m^*_{D,p}$ \cite{baran04,liu02}. When the nonrelativistic mass is 
derived from RMF theory, it shows the same behavior as the 
Dirac mass, namely  $m^*_{NR,n} < m^*_{NR,p}$ \cite{baran04}. 

Relativistic {\it ab initio} calculations based on realistic nucleon-nucleon interactions, 
such as the Dirac-Brueckner-Hartree-Fock (DBHF) approach, are the proper tool 
to answer this question. However, results from DBHF calculations 
are still controversial. They depend strongly  on approximation schemes and 
techniques used to determine the Lorentz 
and the isovector structure of the nucleon self-energy. 

In one approach, originally proposed by Brockmann and Machleidt  \cite{brockmann90} 
- we call it fit method in the following - 
one extracts the scalar and vector 
self-energy components directly from the single particle potential. 
Thus, mean values for the self-energy components are obtained where  
the explicit momentum-dependence has already been averaged out. In 
symmetric nuclear matter this method is relatively reliable but the 
extrapolation to asymmetric matter introduces two new parameters in  
order to fix the isovector dependences of the self-energy components. This  
makes the procedure ambiguous, as has been demonstrated in 
 \cite{schiller01}.  Calculations based on this method 
predict a mass splitting of $m^*_{D,n} > m^*_{D,p}$~\cite{alonso03}. 
On the other hand, the components of the self-energies can directly 
be determined from the projection onto Lorentz invariant
amplitudes. Projection techniques are involved but more accurate and have 
been used e.g. in \cite{thm87,sehn97,gross99}. When 
projection techniques are used in DBHF calculations for asymmetric 
nuclear matter a mass splitting of $m^*_{D,n} < m^*_{D,p}$ is found 
\cite{schiller01,dejong98,vandalen04b}. In the present work we 
compare the Dirac and the nonrelativistic effective 
mass, both derived from the 
DBHF approach based on projection techniques,
in symmetric and asymmetric nuclear matter.  

In the relativistic Brueckner approach nucleons are dressed 
inside nuclear matter as a consequence of their two-body interactions 
with the surrounding particles. 
The in-medium interaction, i.e. the $T$ matrix, is treated in the ladder
approximation of the relativistic Bethe-Salpeter (BS) equation
\beqa
T = V + i \int  V Q G G T,
\label{subsec:SM;eq:BS}
\eeqa
where $V$ is the bare nucleon-nucleon
interaction. The intermediate off-shell nucleons are described by a 
two-body propagator $iG G$.
The Pauli operator Q prevents scattering to occupied states.
The Green's function $G$ which describes the propagation of dressed
nucleons in the medium fulfills the Dyson equation
\beq
G=G_0+G_0\Sigma G.
\label{subsec:SM;eq:Dysoneq}
\eeq
$G_{0}$ denotes the free nucleon propagator whereas the influence of the
surrounding nucleons is expressed by the self-energy $\Sigma$.
In the Brueckner formalism  this self-energy is determined by summing up the
interactions with all the nucleons inside the Fermi sea $F$ in
Hartree-Fock approximation
\beqa
\Sigma = -i \int\limits_{F} (Tr[G T] - GT ).
\label{subsec:SM;eq:HFselfeq1}
\eeqa
The coupled set of
Eqs.~(\ref{subsec:SM;eq:BS})-(\ref{subsec:SM;eq:HFselfeq1})
represents a self-consistency problem and has to be iterated until
convergence is reached. 
The self-energy consists of scalar $\Sigs$ and vector 
$\Sigma^\mu = (\Sigo,  \textbf{k} \,\Sigv) $ components
\beqa
\Sigma(k,\kf)= \Sigs (k,\kf) -\gamma_0 \, \Sigo (k,\kf) +
\bfgamma  \cdot \textbf{k} \,\Sigv (k,\kf).
\label{subsec:SM;eq:self1}
\eeqa
The decomposition of the self-energy into the different 
Lorentz components (\ref{subsec:SM;eq:self1}) requires the knowledge of 
the Lorentz structure of the T-matrix in (3). For this purpose the 
T-matrix has to be projected onto covariant amplitudes. We use the 
subtracted $T$-matrix representation scheme for the projection 
method described in detail in \cite{gross99,vandalen04b}.

The effective Dirac mass is defined as
\beqa
m^*_D(k, \kf) = \frac{M + \Re  \Sigma_s(k, \kf)}{1+ \Re \Sigv (k, \kf)}~~, 
\label{subsec:SM;eq:dirac}
\eeqa
i.e. it accounts for medium effects through the scalar part 
of the self-energy. The correction through the spatial $\Sigv$ 
part is generally small \cite{thm87,gross99,vandalen04b}. 

The effective mass which is usually considered in order to characterize the 
quasi-particle properties of the nucleon within nonrelativistic frameworks 
is defined as
\beqa
m^*_{NR} = |\veck| [dE/d|\veck|]^{-1}~~,
\label{Landau1}
\eeqa
where $E$ is the energy of the quasi-particle and $\veck$ its momentum. When 
evaluated at $k=k_F$ Eq.(\ref{Landau1}) yields the Landau mass related to 
the $f_1$ Landau parameter of a Fermi liquid \cite{baran04,jaminon}.  
In the quasi-particle approximation, i.e. the zero width limit of the in-medium 
spectral function, these two quantities are connected by the dispersion relation
\beqa
E= \frac{\veck^2}{2 M} + \Re U (|\veck|, \kf)~~.
\label{Energy1}
\eeqa
Equations (\ref{Landau1}) and (\ref{Energy1}) yield then the 
following expression for the effective  mass
\beqa
m^*_{NR} = \left[\frac{1}{M} 
+  \frac{1}{|\veck|} \frac{d}{ d |\veck|}\Re U \right]^{-1}~~.
\label{mlandau}
\eeqa
In a relativistic framework $m^*_{NR}$ is obtained from the 
corresponding Schroedinger equivalent single particle potential 
\beq
U (|{\bf k}|,\kf) 
=  \Sigs - \frac{1}{M} \left( E\Sigo - {\bf k}^2\Sigv\right) 
	+ \frac{\Sigs^2 - \Sigma_{\mu}^2}{2M}  ~.
\label{uopt}
\end{equation}
An alternative would be to derive the effective mass from 
Eq. (\ref{Landau1}) via the relativistic single particle 
energy $E = (1+\Re \Sigv)\sqrt{ {\bf k}^{2} + m^{*2}_{D}} - \Re  \Sigo$. 
However, since the  single particle energy contains relativistic corrections 
to the kinetic energy a comparison to nonrelativistic approaches 
should be based on the Schroedinger equivalent potential (\ref{uopt}) 
\cite{jaminon}.

Thus, the nonrelativistic effective  mass is based on a completely different 
physical idea than the Dirac mass, since it parameterizes
the momentum dependence of the single particle potential. Hence, it is 
a measure of the nonlocality of the single particle potential $U$. The 
 nonlocality of $U$ can be due to nonlocalities in space, resulting in 
a momentum dependence, or in time, resulting in an energy dependence. 
In order to clearly separate both effects, one has to distinguish further 
between the so-called k-mass and the E-mass \cite{jaminon}. The  k-mass 
is obtained from eq. (\ref{mlandau}) at {\it fixed} energy while  the  E-mass 
is given by the derivative of $U$ with respect to the energy at {\it fixed} 
momentum. A rigorous  distinction between these two masses requires the knowledge 
of the off-shell behavior of the single particle potential $U$. As discussed e.g. 
by Frick et al. \cite{muether04} the spatial nonlocalities of $U$ are mainly 
generated by exchange Fock terms and the resulting k-mass is a smooth 
function of the momentum. Nonlocalities in time are generated by Brueckner 
ladder correlations due to the scattering to intermediate states which 
are off-shell. These are mainly short-range correlations which generate a 
strong momentum  dependence with a characteristic enhancement of the 
E-mass slightly above the Fermi surface \cite{mahaux85,jaminon,muether04}. 
The effective nonrelativistic mass defined 
by Eqs. (\ref{Landau1}) and (\ref{mlandau}) contains both, nonlocalities in 
space and time and is given by the product of k-mass and E-mass \cite{jaminon}. 
It should therefore show such a  typical peak structure around $\kf$. 
The peak reflects - as a model independent result - the increase of the 
level density due to the vanishing imaginary part of the optical 
potential at $\kf$ which is seen, e.g., in shell model calculations 
\cite{jeukenne76,mahaux85,jaminon}. One has, however, 
to account for correlations beyond mean field or Hartree-Fock 
in order to reproduce this behavior. 

The following results and discussions are based on the Bonn A nucleon-nucleon potential. 
However, they do not strongly depend on the particular choice of the interaction. 
\begin{figure}[!h]
\begin{center}
\includegraphics[width=0.9\textwidth] {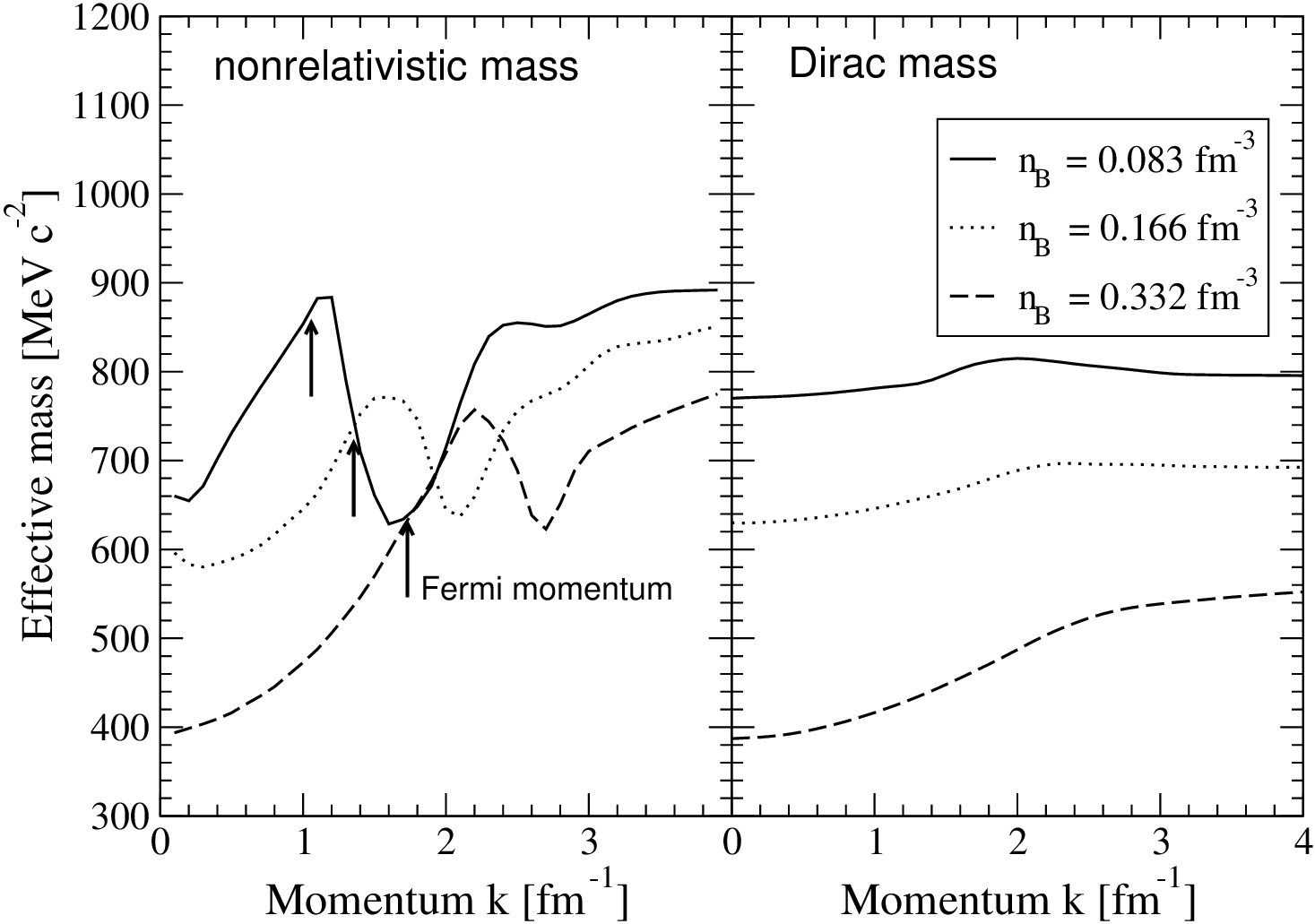}
\caption{The effective mass in isospin symmetric nuclear matter 
as a function of the
momentum $k=|\veck|$ at different densities.
\label{fig:sLD}}
\end{center}
\end{figure}

In Fig.~\ref{fig:sLD} the nonrelativistic effective mass 
and the Dirac mass are 
shown as a function of momentum $k$ at different Fermi momenta 
of $\kf=1.07,~1.35,~{\rm and}~1.7~{\rm fm}^{-1}$ which corresponds to 
nuclear densities $n_B = 4/6\pi^2 \kf^3 = 0.5n_0,~n_0,~{\rm and}~2 n_0$ 
where $n_0=0.166~{\rm fm}^{-3}$ is the 
nuclear saturation density. 
The projection method reproduces a pronounced peak of the 
nonrelativistic mass slightly above $\kf$ as it also seen in nonrelativistic 
BHF calculations \cite{jaminon}. 
With increasing density this peak is shifted to higher momenta and 
slightly broadened. The Dirac mass is a  smooth function of k with 
a moderate momentum dependence. The latter is in agreement
with the `` reference spectrum approximation '' used
in the self-consistency scheme of the DBHF approach~\cite{vandalen04b}. 
Both, Dirac and nonrelativistic mass decrease in average with increasing nuclear 
density. For completeness it should be mentioned that, 
if $m^*_{NR}$  is extracted directly from the 
single particle energy (\ref{Landau1}) instead from the potential 
(\ref{uopt}), results are very similar. Differences occur only   
at high momenta where relativistic corrections to the kinetic energy 
come into play.  

In the relativistic framework the single particle potential and 
the corresponding peak structure of the nonrelativistic mass are the result 
of subtle cancellation effects of the scalar and vector self-energy components. 
This requires 
a very precise method in order to determine variations of the 
self-energies  $\Sigma$ which are small compared to their absolute scale. The 
used projection techniques are the adequate tool for this purpose. 
Less precise methods yield only a small enhancement, i.e. a broad bump 
around $\kf$ \cite{jaminon,thm87}. 
The extraction of mean self-energy components from a fit to 
the single particle potential, is not able to resolve such 
a structure at all. 
\begin{figure}[!h]
\begin{center}
\includegraphics[width=0.9\textwidth] {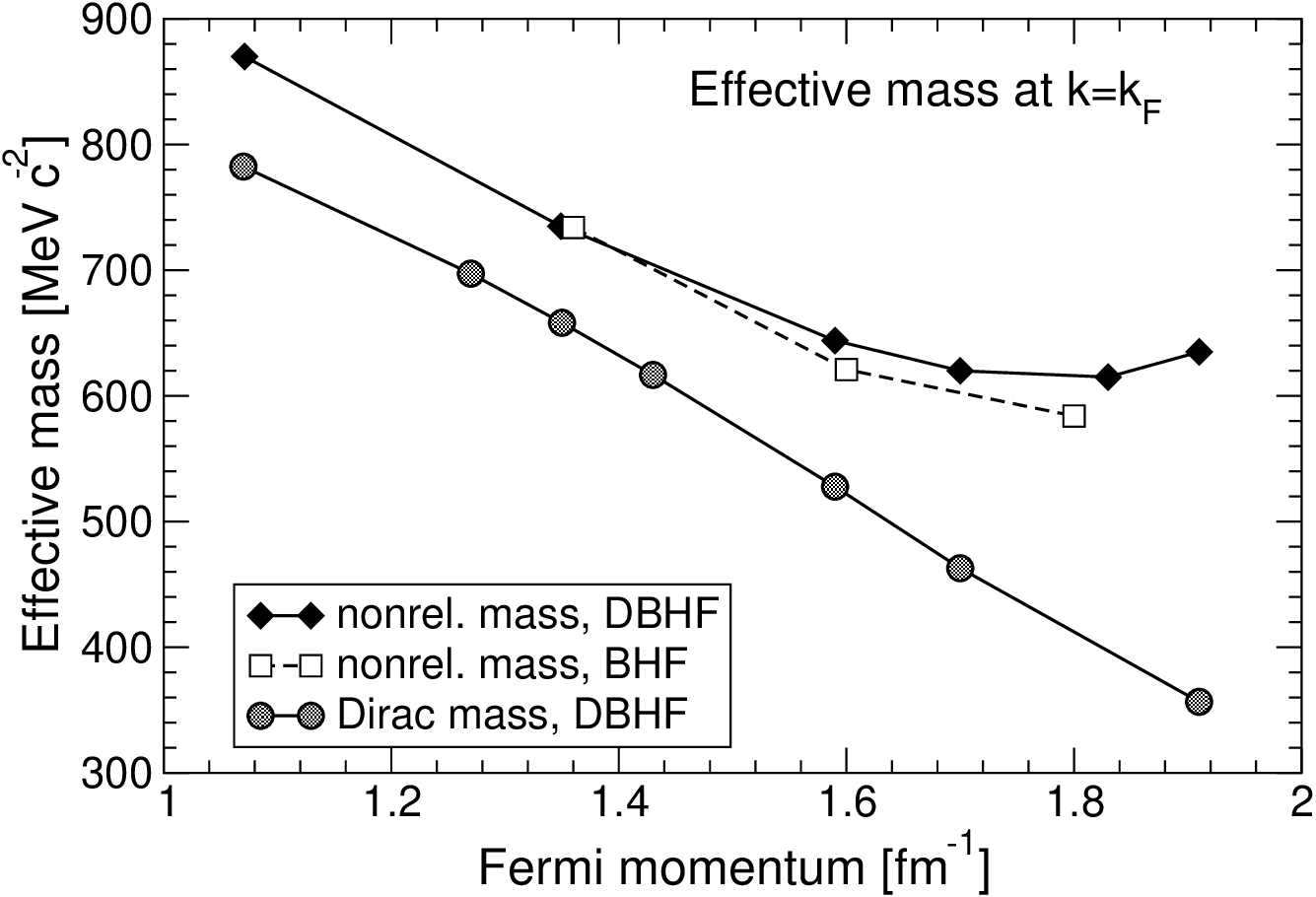}
\caption{The effective mass in isospin symmetric nuclear matter 
at $k=|\veck|=\kf$ as a function of the
Fermi momentum $\kf$.
\label{fig:mstar}}
\end{center}
\end{figure}

Fig.~\ref{fig:mstar} compares the density dependence of the two 
effective masses. Both, the nonrelativistic  and the Dirac mass are determined at the 
Fermi momentum $k=|\veck|=\kf$ and shown as a function of $\kf$.  
Initially, the nonrelativistic  mass decreases with increasing Fermi momentum $\kf$. 
However, at high values of the Fermi momentum $\kf$, it starts to rise again. 
The Dirac mass, in contrast, decreases continously with 
increasing Fermi momentum~$\kf$.  
In addition, also results from nonrelativistic BHF calculations 
\cite{muether}, based on the same  Bonn A  interaction, are shown and 
the agreement between the nonrelativistic and the relativistic 
Brueckner approach is quite good. 
\begin{figure}[!h]
\begin{center}
\includegraphics[width=0.9\textwidth] {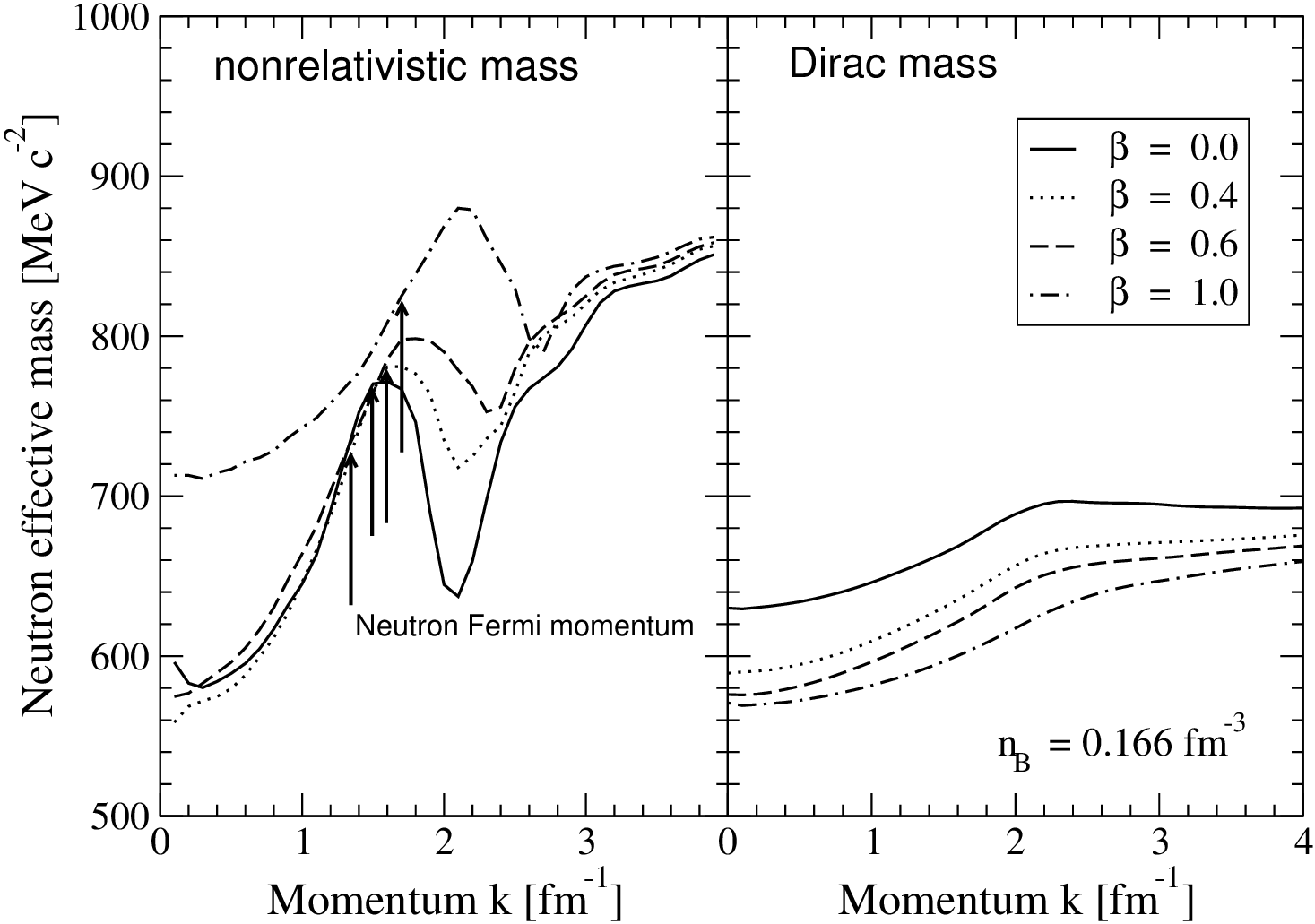}
\caption{Neutron effective mass as a function of the
momentum $k=|\veck|$ for various values
of the asymmetry parameter $\beta$ at fixed nuclear density 
$n_B = 0.166 \quad \textrm{fm}^{-3}$.
\label{fig:aLD}}
\end{center}
\end{figure}

In Fig.~\ref{fig:aLD} the neutron nonrelativistic and Dirac mass 
are plotted for various values
of the asymmetry parameter $\beta = (n_n - n_p)/n_B$ at fixed nuclear density 
$n_B = 0.166 \quad \textrm{fm}^{-3}$. An increase of 
$\beta$ enhances the neutron density and has thus 
for the density of states the same effect as an increase of the density 
in symmetric matter.  Another interesting issue is the proton-neutron mass 
splitting in asymmetric nuclear matter.
Although the Dirac mass derived from the DBHF approach has a proton-neutron mass 
splitting of $m^*_{D,n} <m^*_{D,p}$ as can be seen from Fig.~\ref{fig:aLD}, 
the nonrelativistic mass derived from the DBHF approach 
shows the opposite behavior, i.e. $m^*_{NR,n} > m^*_{NR,p}$ 
which is in agreement with the results from nonrelativistic 
BHF calculations \cite{zuo99,muether04}. In Fig.~\ref{fig:aLD} only 
neutron masses are depicted but the corresponding proton masses always 
behave opposite, i.e. a  neutron mass which is decreasing/increasing  
with asymmetry corresponds to a increasing/decreasing proton mass. 
\begin{figure}[!h]
\begin{center}
\includegraphics[width=0.9\textwidth] {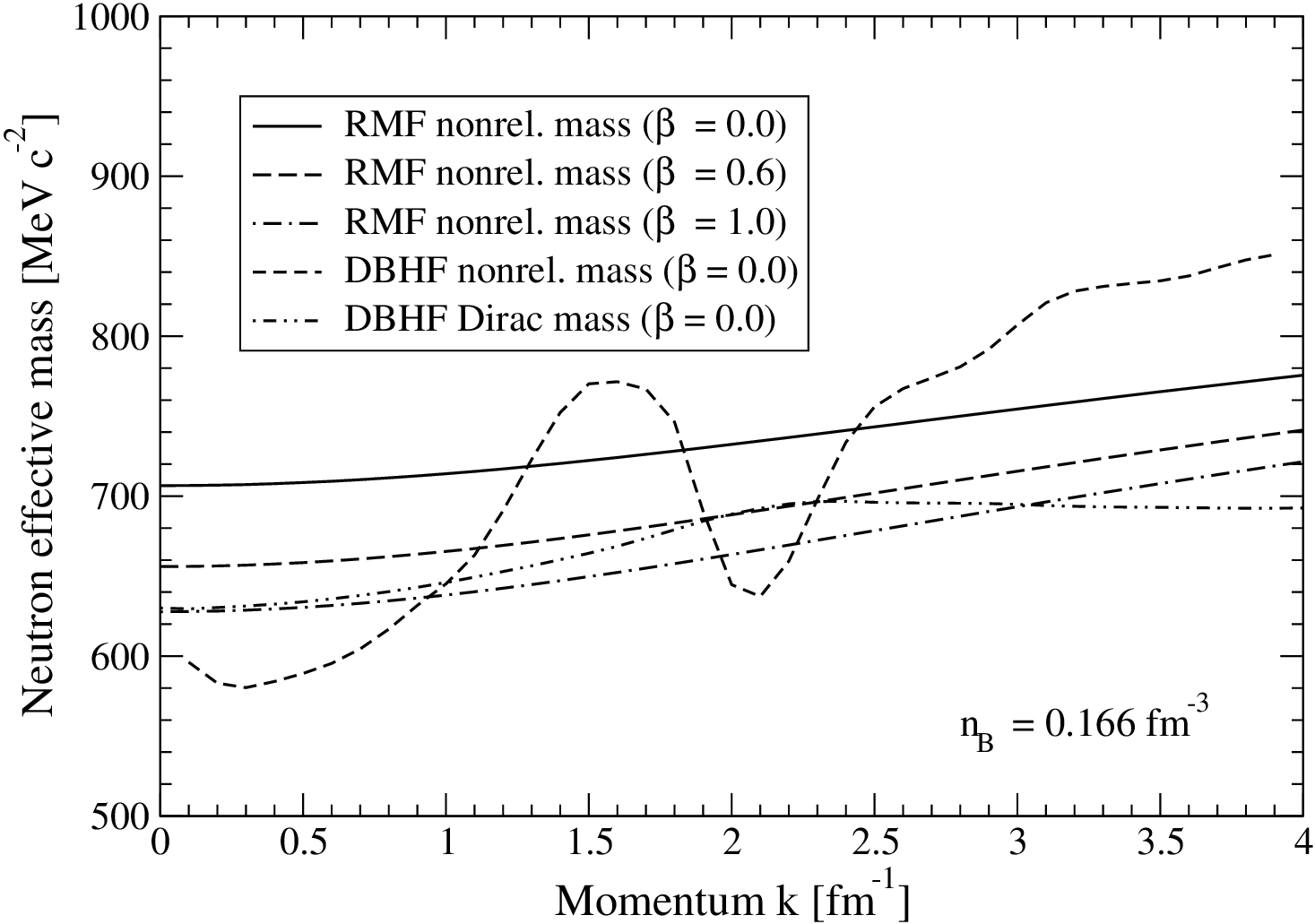}
\caption{Neutron effective mass obtained in the RMF approximation 
as a function of the
momentum $k=|\veck|$ at fixed nuclear density $n_B = 0.166 \quad \textrm{fm}^{-3}$.
\label{fig:RMFasymmass}}
\end{center}
\end{figure}

Fig.~\ref{fig:RMFasymmass} demonstrates finally the influence of the 
explicit momentum dependence of the DBHF self-energy. 
In RMF theory  the Dirac mass and the vector self-energy 
are momentum independent. The nonrelativistic mass is now 
determined from the RMF approximation to the  
single particle potential, i.e. neglecting  
the momentum dependence 
of the scalar  $\Sigs$ and vector fields $\Sigo$ and $\Sigv$ in 
Eqs. (\ref{subsec:SM;eq:dirac}) and   (\ref{uopt}).  
The single particle energy is then given by 
$E_{RMF}= (1+\Re \Sigv(\kf)) \sqrt{|\veck|^2+m^{*2}_D(\kf)}+\Re \Sigo(\kf)$. 
In Fig.~\ref{fig:RMFasymmass} this  'RMF' nonrelativistic mass is 
plotted  for various values
of the asymmetry parameter $\beta$ at $n_B = 0.166 \quad \textrm{fm}^{-3}$.
For comparison  the full DBHF nonrelativistic and Dirac masses for
symmetric nuclear matter are shown as well. Due to the parabolic momentum 
dependence of $E_{RMF}$ the corresponding RMF mass has no bump or 
peak structure but is a continuously rising function with momentum. 
At $k=k_F$ it correponds to the RMF Landau mass \cite{jaminon,matsui81}. 
The RMF nonrelativistic mass decreases with increasing asymmetry parameter. 
RMF theory predicts the same proton-neutron mass splitting for the Dirac 
and the nonrelativistic mass, i.e. $m^*_{D,n} < m^*_{D,p}$ and $m^*_{NR,n} < m^*_{NR,p}$. 
This is a general feature of the RMF approach \cite{baran04}. 
Full DBHF theory is in agreement with the prediction of RMF theory concerning 
the Dirac mass, however, the mass splitting of the nonrelativistic mass is reversed due 
to the momentum dependence of the self-energies, 
respectively the nonlocal structure of the 
single particle potential, which is neglected in RMF theory. 

In summary, effective nucleon masses in isospin symmetric 
and asymmetric nuclear matter have been derived from the DBHF approach based on 
projection techniques. We compared the momentum and isospin 
dependence of the relativistic Dirac mass and the nonrelativistic 
mass which parameterizes the energy dependence of the single particle spectrum. Firstly, the nonrelativistic effective mass 
shows a characteristic peak structure at momenta slightly 
above the Fermi momentum $\kf$ which indicates 
an enhanced level density near the Fermi surface. 
The Dirac mass is a smooth function of k with 
a weak momentum dependence. Secondly, the controversy 
between relativistic and nonrelativistic 
approaches concerning the proton-neutron 
mass splitting in asymmetric nuclear matter has been resolved.
The Dirac mass shows a mass splitting of $m^*_{D,n} <m^*_{D,p}$, in line 
with RMF theory. However, the nonrelativistic mass derived from the DBHF approach 
has a reversed proton-neutron mass splitting $m^*_{NR,n} >m^*_{NR,p}$ 
which is in agreement with the results from nonrelativistic BHF calculations.   

The authors would like to thank B.L. Friman and H. M\"uther 
for helpful discussions.

\end{document}